\documentclass[preprint]{aastex}

\newcommand{\be}{\begin{equation}}
\newcommand{\ee}{\end{equation}}

\begin{document}
\title{On the Radii of Extrasolar Giant Planets} 

\bigskip
\author{Peter Bodenheimer} 
\bigskip
\author{Gregory Laughlin} 
\bigskip
\author{Douglas N. C. Lin} 

\bigskip
\bigskip
\bigskip
\bigskip
\affil{UCO/Lick Observatory, University of California,  Santa Cruz, CA 95064}

\begin{abstract} 
We have computed evolutionary models for extrasolar planets which range 
in mass from 0.1 $M_{\rm JUP}$ to 3.0 $M_{\rm JUP}$, and which range in
equilibrium temperature from 113 K to 2000 K. We present four 
sequences of models, designed to show  the structural
 effects of a solid (20 $M_{\oplus}$)
core and of internal heating due to the conversion of kinetic to thermal
energy at pressures of tens of bars. The model radii at ages of 4--5 Gyr
 are intended for future
comparisons with radii derived from observations of 
transiting extrasolar planets. To provide such comparisons, we expect 
that of order 10 transiting planets with orbital periods less than 200 days 
can be detected 
around bright ($V<10-11$) main-sequence stars for which accurate 
well-sampled radial velocity (RV) measurements can also be readily accumulated.
Through these observations, structural properties of the planets will be
derivable, particularly for low-mass, high-temperature planets. 
Implications regarding the transiting companion to OGLE-TR-56
recently announced by Konacki et al. are discussed.

With regard to the transiting planet, HD 209458b, we find, 
in accordance with other recent calculations, 
that models without internal heating predict a radius 
that is $\sim 0.3 R_{\rm JUP}$ smaller than the observed radius.
Two resolutions have been proposed for this discrepancy.
Guillot \& Showman hypothesize that deposition of kinetic wind energy at 
pressures of tens of bars is responsible for heating the planet and maintaining
its large size. Our models confirm that dissipation of the type proposed
by Guillot \& Showman can indeed produce a large radius for HD 209458b.
Bodenheimer, Lin \& Mardling suggest that HD 209458b
owes its large size to dissipation of energy arising from 
ongoing tidal circularization
of the planetary orbit. This mechanism requires the presence of an
additional planetary companion to continuously force the eccentricity.
We show that residual scatter
in the current RV data set for HD 209458b is consistent with  
the presence of an as-of-yet undetected second companion, and that
further RV monitoring of HD 209458 is indicated.

Tidal circularization theory also can provide constraints on planetary radii. 
Extrasolar giant planets
with periods of order 7 days should be actively circularizing. We find that
the observed eccentricities
of $e \sim 0.14$ for both HD 217107b ($P=6.276~{\rm d}$; $M\sin i =1.80~M_{\rm JUP}$),
and for HD 68988b ($P=7.125~{\rm d}$, $M\sin i =1.29~M_{\rm JUP}$) likely 
indicate either relatively small planetary radii for these objects
($ R \sim 1.1~R_{\rm JUP}$) or tidal quality factors
in the neighborhood of $Q_{\rm P} \sim 10^{7}$. For these two
planets, it will be difficult to differentiate the contribution
from tidal and kinetic heating. But the radius of
HD 168746b ($P=6.403~{\rm d}$, $M\sin i =0.23~M_{\rm JUP}$) is sensitive
to whether the planet's interior is heated by tidal dissipation or
kinetic heating. The tidal circularization time scale of this
planet is shorter than the age of its host star, but we show that within
the observational uncertainties, the published RV data can also
be fit with a circular orbit for this planet.
As more RV planets with periods of
order a week are discovered, $Q_{\rm P} (T_{\rm eq},M_{\rm P})$ and 
$R_{\rm P}(T_{\rm eq},M_{\rm P})$
will become better determined.

\end{abstract}


\section{Introduction} 

Astronomers are understandably enthusiastic about the prospect of detecting Jovian-type 
planets transiting bright (${\rm V}<10$) chromospherically quiet late-type
stars. A bright parent star allows highly accurate orbital parameters to be deduced
from Doppler RV measurements, while the transit phenomenon
affords direct measurements of the planetary parameters.

Indeed, the celebrated occultations of HD 209458 (V=7.65)
(Charbonneau et al. 2000, Henry et al. 2000) have
provided a scientific bonanza. The identification of this transit
allowed detailed follow-up
measurements, including direct
and accurate measurements of the planet's radius 
($1.35 \pm 0.06 R_{\rm JUP}$ [Brown et al. 2001];  $1.41 \pm 0.10 R_{\rm JUP}$ [Cody \& Sasselov 2002]), mass
($0.69 \pm 0.05 M_{\rm JUP}$; [Mazeh et al. 2000]), and even
the presence of sodium in its atmosphere (Charbonneau et al. 2002).

Following the discovery of 51 Peg (Mayor \& Queloz 1995), theoretical models
of Jovian-mass planets subject to strong irradiation were computed (Guillot
et al.  1996). These models predicted that short-period 
Jovian planets with equilibrium 
temperatures of 1300--1400 K at ages of several Gyr 
would be significantly larger than Jupiter.
The discovery that HD 209458b has a large  radius 
initially seemed to be a strong confirmation of these models (Burrows et al.  2000).

While the large observed size of HD 209458b certainly
suggests that it is  a gas-giant composed primarily of hydrogen,
recent work by  
Guillot \& Showman (2002) indicates that a serious gap exists in our understanding of irradiated
giant planets.  They show that standard evolutionary models can recover the
observed radius of HD 209458b only if the deep atmosphere is unrealistically hot.
 A more nearly correct treatment of radiative heating leads to more rapid and efficient
planetary contraction, and predicted radii which are 0.2--0.3 $R_{\rm JUP}$ too small, 
that is, in the range $1.1 ~R_{\rm JUP}$ at an age of 5 Gyr. 
The results of Bodenheimer et al. (2001), based on similar assumptions,  
also produce radii which are too small. 

Two
resolutions to this problem have been suggested. Bodenheimer et al. (2001) argue
that HD 209458b might be receiving interior tidal heating through ongoing
orbital circularization, whereas Guillot \& Showman (2002) propose that
strong insolation-driven
weather patterns on the planet are leading to some conversion of kinetic wind energy into
thermal energy at pressures of tens of bars. At the time the work of Bodenheimer et al.
(2001) was done, the eccentricity of HD 209458 was thought to be consistent with zero,
thus no obvious source of tidal heating was present. However, 
as  we discuss below, the continued accumulation of RV measurements of
HD 209458b will allow for much tighter constraints on the planetary eccentricity
and the possible  presence of a second planetary companion.

Although there are a number of studies of the evolution of giant planets (for 
reviews see Hubbard et al. 2002; Burrows et al. 2001),
at present  there are no published grids of models for planets
with equilibrium temperatures in the 500--1000 K regime that are  intermediate
between Jupiter ($T_{\rm eq} = 113$ K) and HD 209458b ($T_{\rm eq} \approx  1350$ K). As
we discuss below, we expect that transiting planets
with intermediate periods ($10~{\rm d} < P < 200~{\rm d}$) will soon be found.
One of the major intents of this paper is to provide predictions for
the radii of such planets. Indeed, an accurate size and mass determination
for even a single intermediate-period planet will help resolve the size
discrepancy observed for HD 209458b. A planet with intermediate period will
not be experiencing significant internal tidal dissipation, but would still
be irradiated to an extent sufficient to produce a significant amount of
kinetic heating from the mechanism suggested by Guillot \& Showman (2002).
We therefore also wish to show that within certain limits such observations can
(1) distinguish between planets with and without a solid core, and (2) show
whether or not the kinetic heating mechanism is operating. One possible consequence
of such a comparison could be that the kinetic heating mechanism is not effective 
and that therefore HD 209458b's radius can only be explained by tidal dissipation. 
On the other hand, if newly-discovered radii were found to be consistent with
kinetic heating, constraints could be placed on the (still not-well understood)
tidal heating mechanism. 
 
Intermediate-period planets are also interesting because they can harbor
dynamically stable large satellites. Indeed, all satellites
larger than $R=70~{\rm km}$
orbiting a $P=3~{\rm d}$ Jovian planet are removed over 5 Gyr.
However, Mars-mass moons can last for 5 Gyr in the Hill Sphere
of a 1 $M_{\rm JUP}$ planet orbiting a 1 $M_{\odot}$ star
in a 27 day (0.18 {\sc au}) orbit,
whereas in a 54 day (0.28 {\sc au}) orbit, Earth-mass moons are dynamically stable
(Barnes \& O'Brien 2002, Evonuk et al. 2003).
Brown et al. (2001) report that with HST, detections of satellites as small as 1
$R_{\oplus}$
are feasible. Therefore, the discovery of  the transit of an 
intermediate-period planet could
be followed up to search for large moons, and, additionally, planetary rings.

\section{Prospects for Detection of Transiting Planets} 

The {\it a priori}  probability that a planet transits its parent star as seen from the line of
sight to Earth is given by,
$${\cal P}_{\rm transit}=0.0045 \left({ {1 {\rm AU}} \over {a}}\right)\left({R_{\star}+R_{\rm pl}\over{R_{\odot}}}
\right)\left({1-e \cos(\pi/2-\varpi) \over{1-e^{2}}}\right) \,\eqno(1)  $$
where $a$ is the semi-major axis of the orbit, $R_{\star}$ and $R_{\rm pl}$ are 
the radii of the star and planet, respectively, $e$
is the orbital eccentricity, and $\varpi$ is the argument of periastron referenced to the
plane of the sky. In order to obtain an accurate mass for a transiting planet,
one requires a good set of RV measurements so that an orbit can be fit.
Accurate RV measurements are best obtained for bright parent stars, so it is
therefore useful to estimate the number of transiting planets that are likely to be available
for accurate mass and radius determinations.
If one examines the parameters of the current RV planet catalog\footnote{
see, e.g., http://www.transitsearch.org/stardatabase/index.htm }, 
one finds that among the 16 known
planets with periods $P<10~{\rm d}$, there are $<n_{\rm T}>=1.62$ expected transits, and
indeed, within this group, a transiting case (HD 209458b) is known. Twelve of the planets with
$P<10~{\rm d}$ have firm non-detections (including HD 68988b and HD 217107b, whose properties are
discussed in more detail below).
Of the remaining three planets with $P<10~{\rm d}$,
one, HD 76700b ($P=3.971~{\rm d}$)
is currently under surveillance, and two others,
HD 162020b ($P=8.428~{\rm d}$), and HD 168746 ($P=6.403~{\rm d}$)
will be evaluated during Spring 2003.

Within the aggregate of 23 known  planets having periods in the range $10~{\rm d}<P<200~{\rm d}$, the expected
number of transiting planets is $<n_{\rm T}>=0.64$. Very few of the parent stars in this group, however,
have been exhaustively monitored for transits. These stars therefore represent excellent targets
for a distributed network of small telescopes.
We also note that among the 60 known planets with $P>200~{\rm d}$, one expects $<n_{\rm T}>\approx 0.5$
additional transiting planets. Because of the less well-determined orbits for most of these
long-period planets, and because of the infrequent occultations, these stars will remain
difficult targets to follow up.

In addition to the current census, more planets with periods suitable
for the discovery of transits
will soon be emerging from the RV surveys. To
see this, consider Figure 1, which plots the periods of 98 known extrasolar planets versus
the Julian dates on which their discoveries were announced. The magnitudes of the parent stars
are color-coded to range from red ($V<4$) to black ($V>9$), while the radii of the circles
marking each planet are proportional to $(M_{\rm P} \sin i)^{1/3}$. This diagram shows that the
pace of discovery of extrasolar planets with $P<200~{\rm d}$ is proceeding at a steady rate,
while the rate of discovery of planets with $P<10~{\rm d}$ has begun to peak. Currently, within the
$10~{\rm d}<P<200~{\rm d}$ range, there are five known planets orbiting stars with $V<6$, seven orbiting stars
with $6<V<7$, six orbiting stars with $7<V<8$, and two planets each in the $8<V<9$, and $9<V<10$
ranges. If we assume that every available chromospherically quiet main sequence dwarf with
$V<6$ has been adequately surveyed for $P<200~{\rm d}$ planets, and that each magnitude bin
of unit width
contains 1.8 times as many stars as available for bin $(V-1)$, then we expect that roughly
$9+16+29+52=106$ detectable planets with $10~{\rm d} < P < 200~{\rm d}$ exist in orbit around stars
with $V<10$, indicating that more than 100 additional planets in this category can be detected
using current RV techniques for bright stars. Statistically, this implies that
3 intermediate-period RV-detectable transiting planets orbit bright nearby stars.
Application of the same argument to planets with $ P < 10 {\rm d}$ indicates
that 4 short-period transiting planets are to be expected in orbit around bright stars.

In summary, we can expect to obtain highly accurate radii and 
masses for only a limited number of planets. This collection will serve as the major 
observational basis for our understanding of the structure of giant planets. The ten
or so transiting planets that we can expect to find will likely span two orders of 
magnitude in mass ($0.1 M_{\rm JUP} < M_{P} < 10.0 M_{\rm JUP}$), and a wide 
range in temperature ($300~{\rm K}<T_{\rm eq} <2000~{\rm K}$). In order to interpret this data, 
it is useful to compute a grid of models spanning  mass  and equilibrium temperature at the
age of a typical planet-bearing star (5 Gyr). 

\section{Models for Irradiated Giant Planets} 

Models for the evolution of giant planets of various masses
have been computed with a descendant of the Berkeley stellar evolution code
(Henyey, Forbes, \& Gould 1964). This computer program has received
a number of modifications and improvements to the input physics during
its forty years of existence. It calculates the evolution and mass accretion
rate of the gaseous envelope, under the assumption that the planet is 
spherical, and that the standard equations of stellar structure apply.
It has been used to calculate
the formation phase of planets in the solar system (Pollack et al. 1996) 
and of extrasolar planets by
Bodenheimer, Hubickyj, \& Lissauer (2000). 
The physical assumptions employed in the calculations are
described in those papers.  Energy transport  occurs either by  radiation or
convection, according to the
Schwarzschild criterion for convection. Energy sources include (1) gravitational
contraction, (2) cooling
of the interior, and (3) in some cases energy deposition in the atmosphere, at
pressures in the range
1 bar to 100 bars,  caused by stellar heating (Guillot \& Showman 2002).
In the outer envelope of the planet, where radiation is likely to be the energy
transport mechanism,
dust grains are assumed to have settled into the interior and evaporated, so
that they contribute negligibly to the opacity.  The assumed absence of grains in
the envelopes of low-mass objects is consistent 
with studies of the detailed observed
spectra of T-dwarfs (Tsuji 2002). Pure molecular opacities 
in the temperature range 70-4000 K
were obtained from R. Freedman (private communication) 
for a near-solar composition.
Above 4000 K the table from Alexander \& Ferguson (1994) was used; 
however for such 
temperatures the models are always convective, so the details of the opacity are
unimportant. In convection zones, which always 
include most of the mass of the planet,
the temperature gradient is assumed to take the adiabatic value.  The equation
of state of Saumon, Chabrier, \& van Horn (1995) was employed, 
slightly softened as 
suggested by Burrows et al. (2000). At the surface, 
the luminosity is composed of two components:
the internal luminosity generated by the planet ($L_{\rm int}$), and the energy
absorbed from the stellar
radiation flux and re-radiated (`insolation'). The boundary conditions at the
Rosseland mean
photosphere are
$$
L_{\rm tot} = L_{\rm int} + 4 \pi R^2_{\rm P} \sigma T^4_{\rm eq} \eqno(2)
$$
$$
\bar \kappa_{\rm ph} P_{\rm ph} = {{2} \over {3}} g  \eqno(3)
$$
where $L_{\rm tot}$ is the total luminosity, $T_{\rm eq}$ is the equilibrium
temperature defined
below in equation (8), $\bar \kappa_{\rm ph} $ is the mean opacity at the photosphere,
$P_{\rm ph}$ is the
pressure at the photosphere, and $g$ is the surface gravity. Since the details of
the atmosphere, including frequency-dependent
effects (Chabier \& Baraffe 2000), are not taken into
account, the radii derived by this procedure should be regarded as preliminary.

In the case of extrasolar planets, it is not known whether or not a central
solid/liquid rock/ice core
is present; however its presence or absence could be an important clue to the 
formation mechanism.
The composition, density, and mass of the core would depend on the formation and
migration history of
the planet.
In the case of the solar system giant planets it is likely that cores exist; 
however, the available constraints from
observations and modeling still allow a considerable range in their properties
(Wuchterl, Guillot, \& Lissauer 2000). In the case of Jupiter, the range of 
possible core masses is 0 to 10 M$_\oplus$,
and for Saturn it is 5--15 M$_\oplus$. However these planets also have heavy
element abundances in excess of solar abundance ratios in their gaseous
envelopes. The total heavy element abundance in Jupiter is thought
to be about 30  M$_\oplus$, or 10\% of the total mass,
while in Saturn it is thought to be about 25\% of the total mass. 
The division of the heavy elements
between envelope and core is uncertain mainly because of uncertainties in the
Saumon-Chabrier-van Horn equation of state.
With these considerations in mind, we made calculations for each mass and 
temperature both with a core and without a core. The core mass, which is 
designed to represent approximately the total excess in heavy elements
over solar abundance, is taken to be 40 M$_\oplus$ in the case of the more
massive planets (0.69 M$_{\rm JUP}$ and above) and 20 M$_\oplus$
for lower masses. Two cases with a 20 M$_\oplus$ core for 0.69 M$_{\rm JUP}$ 
were also calculated. Models without cores are assumed to have solar 
composition ($X=.70$, $Y=.28$, $Z=.02$). In the models with cores, the core
density is assumed to have a constant value of 5.5 g cm$^{-3}$,
except for the highest mass
($3 M_{\rm JUP}$), where the high pressures 
in the center require a higher core density, which we 
take to be 10.5 g cm$^{-3}$. The composition of the envelopes of the models with
cores is assumed to be solar. For the case of 1 M$_{\rm JUP}$,
calculations were also made with a core density of
10.5 g cm$^{-3}$; in an example of a current Jupiter model, 
Marley (1999) indicates that the range of densities in the core
is about 9 to 23 g cm$^{-3}$. However, much of the heavy-element 
material in Jupiter is in the envelope at much lower pressures. 
For the lower masses, a core density
of 5.5 g cm$^{-3}$ is thought to be reasonable; 
Marley's (1999) Saturn model has a density of 
6--7 g cm$^{-3}$ in much of the core. A giant planet with enhanced heavy-element
abundance can have a significantly smaller radius than one of the 
same mass with solar abundance.

The initial condition
is a planet of $R\approx 2~R_{\rm JUP}$ at an age of a few Myr. 
The planet is assumed to have
migrated to its final orbit during the formation phase,
thus the assumed equilibrium 
temperature is constant in time.  A calibration
run with $1~M_{\rm JUP}$ with a core and with
insolation appropriate to Jupiter at 5 AU ($T_{\rm eq} = 113$ K)
contracted to a final radius of $1~R_{\rm JUP}$  after an
evolution time of 4.5 Gyr. At the final time, the temperature 
at 1 bar pressure in the atmosphere
atmosphere was about 160 K, in good agreement with observations of Jupiter.

For each value assumed for the planet mass 
(0.11, 0.23, 0.69, 1.00, and $3.0~M_{\rm JUP}$), the
models in Table 1 illustrate how the presence of a solid core, 
as well as the assumed
value of $T_{\rm eq} $, affects the planetary radius (given in R$_{\rm JUP}$). 
In this table, as well as in Table 2,
the core densities are 5.5 g cm$^{-3}$ for all cases except $3.0~M_{\rm JUP}$, 
where it is 10.5 g cm$^{-3}$.
Table 2 presents an exactly analogous sequence of models which also include
an energy source term  (in addition to the direct 
radiative heating of the atmosphere),
applied as described by Guillot \& Showman (2002). This
source term deposits 1.7\% of the incoming stellar flux in the
regions of the planetary envelope where the pressure is tens of bars. 
The energy deposition
fraction of  1.7\% was obtained through calibration runs
in which the fraction was varied 
until the radius of HD 209458 was  reproduced. Specifically, with that fraction,
for the case of 0.69 M$_{\rm JUP}$, without a core and with $T_{\rm eq}
= 1300 K$, a radius of  1.37 R$_{\rm JUP}$ was obtained at an age of 4.5 Gyr.
For all other models, the same fraction of the incident stellar radiation 
was assumed to be
converted to heat at pressure levels from the 
surface down to $\sim 100$ bar, with an
exponential fall-off of heating with depth. The incident stellar radiation is
of course proportional to the projected area of the planet's 
star-facing hemisphere; this
effect was taken into account during the contraction.

The radii of these models show in many cases  observably large variations as the
underlying physical parameters (i.e. presence of a core
or energy deposition via kinetic heating) are varied. It is therefore
likely that the discovery of several more transiting extrasolar planets will
make it clear whether giant planets have solid cores, and whether they
generally have access to an interior energy source such as the kinetic heating
described by Guillot \& Showman (2002). We note that the delineation 
between individual modes is especially clear 
for both the hotter and the lower-mass planets. For a low-mass
relatively short-period planet such as HD 168746b with $M \sin i =0.23~M_{\rm JUP}$,
$T_{\rm eq}=1000$ K, and P = 6.4 days, the two models 
without kinetic heating have final radii of 0.90 and 1.07
R$_{\rm JUP}$ for cases with a core and without a core, respectively,
and the two corresponding models with
kinetic heating have radii of 1.34 and 1.53 R$_{\rm JUP}$, respectively.
The four possibilities should be cleanly separable observationally, as the
observational uncertainly is about 0.1 R$_{\rm JUP}$. 
In the case of a planet with 0.11 M$_{\rm JUP}$ (cf. HD 49674),
the effect becomes even larger, with observationally separable
differences among the four cases out to $T_{\rm eq} = 500$ K (about 0.4 AU
for a solar-type star). In the case with a core, 
this model is intermediate between a Saturn-type
and a Neptune-type planet, with an envelope mass of 15 M$_\oplus$ 
and a core mass of 20 M$_\oplus$.

It is important to note, however, that
for a planet as massive as 3 M$_{\rm JUP}$ it is not possible to
observationally distinguish between these various cases,
except that at very short periods with $T_{\rm eq} = 1500$ or 2000 K, 
it may be possible to determine whether or not kinetic heating is
occurring. For 1 M$_{\rm JUP}$ it is in general not possible to distinguish
between planets with or without cores; for the calculations 
without kinetic heating the difference is only about 5\%. It is, however,
possible to distinguish between planets with or without 
kinetic heating if the equilibrium temperature is above about 1200 K. 
Models with 1 M$_{\rm JUP}$ have also been calculated with a core
density of 10.5 g cm$^{-3}$. Their radii are close to 3\%
smaller in all cases (with or without kinetic heating) 
than in the case with a density of 5.5 g cm$^{-3}$,
making the difference in radius between a planet 
with and without a core marginally detectable in some cases.
In the case of 0.69 M$_{\rm JUP}$, planets with or without 
cores are only marginally separable, while planets with or without kinetic
heating are cleanly separable down to $T_{\rm eq}=1000$ K.
Calculations were also made for  0.69 M$_{\rm JUP}$ with 
a core of 20 M$_\oplus$, with and without kinetic heating
and with  $T_{\rm eq} = 1000$ K. The radii turned out to be 1.06 and
1.22  R$_{\rm JUP}$, respectively, intermediate between the values for 
the case without a core and for the case with a core of 
40 M$_\oplus$. In this situation the presence of a core could not be
detected, but the effect of kinetic heating could still be clearly discerned.

\section{Constraints Provided by the Current Planetary Census}

While additional transiting planets
will certainly be of immense importance in clarifying
the structural theory of extrasolar giant planets, there is interesting
information which can be gleaned from the current census of extrasolar 
planets. We now use the planet models which were described in the previous section to 
examine a number of interesting individual cases.

\subsection{HD 209458b}

As Table 1 shows, a standard model for $M=0.69\, M_{\rm JUP}$ without a core at $T_{\rm eq}=1300 {\rm K}$
has a radius of only 1.12 $R_{\rm JUP}$; the model with a core is even smaller.
Bodenheimer, Lin \&  Mardling (2001) suggest that internal dissipation of tidal
energy arising from orbital circularization is responsible for the large 
observed size of HD 209458b. If HD 209458b is unperturbed by a third body, then
tidal dissipation will circularize its orbit on a timescale (Goldreich \& Soter 1966)
$$\tau_{\rm circ}={e\over{\dot{e}}}=\left( {4Q_{\rm P}\over{63n}}\right)
\left( {M_{\rm P}\over M_{\star}}\right)\left( {{a} \over{ R_{\rm p}}}\right)^5
 =0.082 \left( {Q_{\rm P}\over{10^{6}}} \right)
{\rm Gyr}\, ,\eqno(4)$$
where $n=2\pi/P=1.783~{\rm rad}/{\rm d}$ is the planet's mean motion, 
$M_{\star}=1.1 M_{\odot}$ is the stellar mass, and $Q_{\rm P}$ is the tidal
quality factor. $Q_{\rm P}$ is associated with substantial uncertainty. Based on
the Jupiter-Io interaction, $Q_{\rm JUP}$ is estimated to lie in the range
$6\times10^{4}-2\times10^{6}$ (Yoder \& Peale 1981). Onset of tidal circularization 
among main-sequence binaries suggests that the Q-value for stars
is of order $10^{6}$ (Terquem et al. 1998). The $Q_{\rm P}$ values for
extrasolar planets are likely to be strong functions of planetary mass,
temperature, and composition. As more planets are found with periods,
$P\sim{7 \rm d}$, the range of $Q_{\rm P}$ values appropriate to moderately
irradiated planets of a range of masses will become better determined.

The rate of internal energy dissipation is given by,
$${\dot{E}}_{d}={e^{2}GM_{\star}M_{\rm P} \over{a \tau_{\rm circ}}}, \,\eqno(5) $$
which for HD 209458b is $$\dot{E}_{d}= 1 \times 10^{29} e^{2}\left({10^{6}\over{Q_{\rm P}} }\right)  
~~ {\rm erg}~{\rm s}^{-1}\, .\eqno(6) $$
To estimate this rate for particular choices of $Q_{\rm P}$, we need to estimate the orbital eccentricity for
HD 209458b.

To date, the UC-Keck Planet Search has obtained RV
measurements of HD 209458 which fall in 33 distinct 2-hour time bins. These
updated velocities have been provided by G. Marcy (2002, personal communication).
Among these 33 binned observations, 3 have photon-weighted epochs which fall
within the periodic transit window for HD 209458b. Radial velocity measurements
taken during transit are seriously affected by asymmetric distortions in the stellar
line profiles arising from the planet occulting a rotating star
(Bundy \& Marcy 2000, Queloz et al. 2000). We therefore remove these three points
from the RV data set and use a Levenberg-Marquardt algorithm
(Press et al.  1992) to fit a Keplerian orbit to the remaining velocities.
Repeated observations of the HD 209458b transits which include mid-1990's data
from the Hipparcos epoch photometry (Castellano et al. 2000, Robinchon \& Arenou
2000), HST/STIS observations by Brown et al. (2001), and HST/FGS observations
by Welsh et al. (2003) have allowed the planetary period and the transit
midpoint to be determined to high accuracy. Welsh et al. (2003) report a period
$P=3.5247542 \pm 4.4\times10^{-6} \, {\rm HJD}$ and a 
transit midpoint $T_{c}=2452223.89617 \pm 8.6 \times10^{-5} \, {\rm HJD}$. With
this ephemeris,
and for any given choice of the planetary $e$ and $\varpi$, we can compute 
the mean anomaly $M$ at the epoch of the first data point (JD=2451341.120).
We fit for (1) the planetary eccentricity $e$, (2) the argument of
perihelion $\varpi$, (3) the planetary $M\sin i $, and (4) the
velocity zero-point for the data set.
This four-parameter variation returns
a best-fit system having $e=0.033$, $\varpi=67.17$, $M\sin i =0.679~M_{\rm JUP}$,
and $\Delta v=5.36~{\rm m/s}$. The fit has a $\sqrt {\chi^{2}}=1.69$,
and an rms scatter of $8.05~{\rm m/s}$. If the eccentricity for the planet
is forced to be zero and the velocities are re-fit for $M$,
$M\sin i $, and $\Delta v_{1}$, we obtain
a best fit having $M\sin i =0.652$, and $\Delta v_{1}=4.74$ with an rms
scatter of $8.31~{\rm m/s}$. These results are in excellent agreement with
those obtained by Marcy (2002; personal communication) who finds a
best-fit eccentricity $e=0.028 \pm 0.012$ using an independent code.

The measured eccentricity of $e \sim 0.03$ in the one-planet fit to the
HD 209458b RV data therefore appears to have some statistical significance.
For $e=0.03$, the rate of energy dissipation in HD 209458b is ${\dot E}=
1.0 \times 10^{26} ({10^{6}\over{Q_{\rm P}}})~{\rm erg~s^{-1}}$. A new model calculation was made for the
evolution of 0.69 M$_{\rm JUP}$ at $T_{\rm eq} = 1300 $ K including tidal dissipation energy
distributed uniformly in mass throughout the gaseous region of the planet. It was determined that 
 a no-core model requires internal tidal 
heating of  ${\dot E}_{\rm d}\approx 4 \times 10^{26}$ erg s$^{-1}$ to achieve a radius $R=1.35~
R_{\rm JUP}$. Therefore, if the best-fit eccentricity is secure, and if HD 209458b
contains no core, the observed radius can be explained by tidal dissipation if $Q_{\rm P}  \approx 2.5 
\times 10^5$, well within the range derived for Jupiter. The required dissipation is higher than
that quoted by Bodenheimer et al. (2001) because of changes in the equation of state and
opacity since that time. 
Also, for a model with a core, internal heating of $\approx 4\times10^{27}
~{\rm erg~s^{-1}}$ is required to maintain the same radius. 
This amount of heating would require, for the same value of $Q_{\rm P}$, 
an eccentricity $e \approx 0.1$, a value which appears to be incompatible
with the RV data set.

Any non-zero eccentricity for HD 209458b implies that some  mechanism exists
to excite eccentricity, since the circularization
e-folding time ${e\over{\dot{e}}}\sim 10^{8}$ years
is considerably shorter than the estimated system age of 4.5 Gyr (Mazeh et al. 2000).
Bodenheimer et al. (2001) suggest that the eccentricity 
of HD 209458b could be forced by an 
additional planetary companion, and this
process does seem to be occurring in the multiple system $\upsilon$ And. 
 Here we examine whether the residual
scatter in the HD 209458 RV data set can admit a second planet
which is capable of forcing a time-averaged eccentricity $\bar{e}\sim0.03$ for HD 209458b.

Figure 2 shows a Lomb-Scargle (Press et al. 1992) periodogram of the RV residuals 
after the RV contribution
due to our 1-planet (e=0.033) fit has been subtracted. The Lomb-Scargle periodogram is
optimized to detect periodicities in unevenly sampled data, and is described
in detail by Scargle (1982). In our periodogram, modest
peaks remain at 365 and 80 day periodicities. The 365 day periodicity is certainly an aliasing peak 
arising from the 1-year observing cycle. If we make the hypothesis that the 80 day peak arises
from a second companion, we estimate the $M\sin i $ of this second planet to be 
$\sim0.12~M_{\rm JUP}$, in order to model the $6.0~{\rm m/s}$ scatter which is unaccounted
for by the rms instrumental error of  $4.8~{\rm m/s}$ and an assumed stellar jitter of $<j>=4.0~{\rm m/s}$.
This estimate for stellar jitter is based on values reported for several similarly old, chromospherically
quiet G stars (Saar, Butler, \& Marcy 1998), and could be significantly larger.

We have performed four  self-consistent 2-planet fits to the 
velocities (see Laughlin \& Chambers 2001). In the first three fits,
we constrain the parameters of the hypothetical second planet ``c''
by fixing the argument of periastron $\varpi_{c}=60.5^{o}$, and
$M_{c}\sin(i)=0.127 M_{\rm JUP}$, and assuming eccentricity values
(i) $e_{c}=0.0$, (ii) $e_{c}=0.2$, and (iii) $e_{c}=0.4$. The parameters
of the inner planet ``b'' are allowed to vary as explained above. The
mean anomaly at the epoch of the first data point and the period
of the hypothetical planet ``c'' are allowed to vary. These fits
are listed as fits(1-3) in Table 3. In each case, the
fitted period is $P_{c}\sim \, 84 \,{\rm d}$. We find that all three fits
can self consistently reduce the excess scatter to that expected from
instrumental noise, added in quadrature to  a stellar jitter of
$<j>=4.1\, {\rm m/s}$ for $e_{c}=0.4$, $<j>=4.2 \,{\rm m/s}$ for $e_{c}=0.2$,
and $<j>=4.3 \, {\rm m/s}$ for $e_{c}=0.0$. We also performed a fit to the RV data in
which we allowed all of the parameters of the hypothetical planet ``c''
to vary, including the mass and the eccentricity. The resulting system
produced an extremely close fit to the data, requiring $<j>=1.7 \, {\rm m/s}$,
which is almost certainly smaller than the velocity jitter of the star.
In this fourth model, listed as fit 4 in Table 3, the
eccentricity of ``c'' is 0.7, and the mass has increased to $0.22 \, M_{\rm JUP}$.
It is important to stress that these fits illustrate only the existence
of models that provide a consistent explanation of the stellar RV variations.
They by no means constitute the detection of a second planetary companion.

We next examine the dynamical consequences arising from the presence of 
a second planet in the HD 209458 system. For concreteness, we use
our fit in which the eccentricity of the hypothetical
second planet is fixed at $e_{c}=0.4$.
Inclusion of this second planet causes the fitted eccentricity of 
HD 209458b to decline to $e_{b}=0.019$. 
Fischer et al. (2001) have noted that the inclusion
of a second planet into a RV fit generally causes the fitted eccentricity
of the first planet to decline. 

In order to examine the degree of eccentricity
exchange between planet ``b'' and the hypothetical planet ``c'',  we start 
with the osculating orbital elements reported in Table 3 for the trial
self-consistent 2-planet fit having $e_{c}=0.4$, and integrate the 
system forward in time using the Burlirsch-Stoer method.

We note that in the absence of 
relativistic advance of the perihelion of planet HD 209458b, the Laplace-Lagrange
secular exchange of angular momentum (e.g. Murray \& Dermott 1999) is quite 
strong -- the eccentricity of planet ``b'' oscillates
between 0.018 and 0.14 with a period of $P\sim40,000$ years. 
Relativistic apse precession, however,
detunes the secular exchange and reduces the time-averaged $\bar{e}$ to 0.03
(see also Mardling \& Lin 2002). The results are shown in Figure 3. The
two planets experience an apsidal lock when relativistic apse precession
is included, but this lock is not required for the secular eccentricity
exchange to occur. Note that while the ratio of the total angular momenta
$L_{b}/L_{c}\sim1.6$ is not far from unity, the larger eccentricity excursions
for planet ``b'' occur because its eccentricity is smaller. That is, given
that the orbital angular momentum of a planet is
$$L_{\rm P}={M_{\rm P}M_{\star}\over{M_{\rm P}+M_{\star}}} a^{2} n {(1-e^2)}^{1/2}\, ,\eqno(7)$$
and that the total angular momentum $L=L_{b}+L_{c}$ is conserved, the $(1-e^{2})^{1/2}$
dependence in the above equation demands larger eccentricity variations
for the planet with smaller eccentricity.

As discussed above, our evolutionary models show that $\dot{E}$ for $\bar{e}=0.03$
produces sufficient heating in the planet to account 
for the observed radius in the event that the planet contains no core and 
$Q_{\rm P}\approx 2.5 \times 10^{5}$. 

The rms RV measurement errors for
HD 209458b are $<\sigma>=4.86~{\rm m/s}$, and so as more RVs are obtained for
this star, the presence of a second companion of the type described here should be confirmed or 
ruled out.
The potential detectability of the perturbing companion 
``c'' is aided by the requirement that
its period should be less than the current $\sim 1200$ 
day duration of the RV data set. HD 209458 shows
a RV trend of only $0.0007 \pm 0.002~{\rm m s^{-1} d^{-1}}$. 
Trial calculations of secular eccentricity exchange indicate that an exterior planet producing
a linear $0.0007~{\rm m s^{-1} d^{-1}}$ velocity trend
is not large enough
to maintain a significant time-averaged eccentricity for planet HD 209458b.
 Hence the perturbing 
companion, if it exists, has a period short enough to average out a residual trend over a 
4-year time frame. The hypothetical $P=80~{\rm d}$ planets considered above easily 
fulfill this condition.

The current Keck Planet Search RV data set for HD 209458b contains only  
33 binned RVs, which means
that a significant improvement of the system characterization
can be obtained over the next several years if frequent additional
measurements are made. 
Because individual high-precision RV measurements
are expensive, it is therefore useful to examine how the detectability
of a hypothetical HD 209458 ``c'' 
improves as more RV measurements are taken.

To do this, we have taken the hypothetical 2-planet system shown in fit 3 of Table 3, and
integrated it forward through JD 2453266 (September 2004). We then generated
a simulated campaign of 120 additional RV measurements over the
next 1.5 observing seasons for HD 209458 (RA=22:03, DEC=+18:53). Our simulated
campaign includes 11 individual RV measurements cadenced according to typical observing
runs on the Keck telescope and 109 RV measurements accumulated over 6
dedicated nights on a smaller instrument such as the Lick Observatory
3-meter telescope. This combination of intensive and sporadic monitoring
is intended to improve phase coverage of both planets, while expending
a reasonable amount of telescope time. 

We then sample the integrated reflex velocity of the star in response to
the two planets at each of the 120 epochs in the simulated observing campaign.
Gaussian scatter consistent with instrumental error (4.8 m/s) and
stellar jitter (4.0 m/s) is added to the sampled stellar reflex velocities.
This results in a synthetic data set which includes the 30 real plus 120 simulated
RV points. We then perform 1-planet and 2-planet fits to all 150 points.
The best 2-planet fit results in a total rms scatter of 6.79 m/s,
while the best 1-planet fit results in a scatter of 7.41 m/s,
suggesting  that a moderately intensive campaign on HD 209458b will 
result in a slow, but nevertheless feasible discrimination between the
1-- and 2--planet hypotheses. Additional Monte-Carlo simulations of this type can
be used to further refine the observing strategy. If the eccentricity of the
hypothetical planet ``c'' is larger, as in the fourth fit of Table 3, then
the planet will be easier to detect, due to the comparatively larger reflex
velocity of the star during the peri-astron passage.

We also note that the time $\Delta T_{c}$ between successive transits for HD 209458b
varies by $\pm 3 {\rm s}$
over the course of a single 84 day period of the hypothetical companion. While the
average
period of HD 209458b can be determined very accurately by spanning a large number
of
periods (Welsh et al. 2003), it is difficult to obtain 1 s accuracy for the duration
of a single orbit. It might be possible, however, to measure precession-induced
secular changes in the period of HD 209458b arising from perturbations caused
by a second companion (see Miralda-Escud\'e 2002).

\subsection{Ongoing Circularization}

The period range between 3 and 10 days is now populated by 16
planets covering a wide range of masses and equilibrium temperatures.
Equation  (4) indicates that if $Q_{\rm P} \sim 10^{6}$, then planets
with periods of order 1 week should be in the process of actively
circularizing. The circularization timescale, however, depends
sensitively on the planetary radius ($\tau_{\rm circ} \propto
R_{\rm P}^{-5}$), and so among planets with periods of about a
week there should be a wide range of non-zero eccentricities,
even in the absence of companions capable of forcing eccentricity.
There are two additional uncertainties. Tidal dissipation within
the host star can also contribute to the evolution of the planet's
eccentricity, although it is smaller than that due to planetary 
dissipation (Dobbs-Dixon et al 2003). Past tidal inflation 
instability can also lead to the circularization of planets with
$P<10 {\rm d}$ (Gu et al. 2003). This effect is not important for
systems with finite eccentricity.
As more planets are discovered in the $3~{\rm d} < P < 10~{\rm d}$
period range, all of the parameters in equation  (4) will begin to
become overconstrained, eventually allowing $Q_{\rm P}(T_{\rm eq}, M_{\rm P})$ and
$R_{\rm P}(T_{\rm eq}, M_{\rm P})$ to be observationally determined.

In order to see how the discovery of additional planets with periods
of order 1 week will lead to  a better determination for $Q_{\rm P}$
as a function of planetary mass and temperature,
consider Table 4, which lists relevant planetary and stellar
information for HD 68988b (Vogt et al. 2002), HD 168746b (Pepe et
al. 2002), and HD 217107 (Fischer et al. 1999). In this table, stellar
masses, radii and effective temperatures are taken from Allende
Prieto \& Lambert (1999).

The equilibrium temperature of the planets is calculated as
$$T_{\rm eq} = { \left[ {(1-A)L_{\star} \over{ 16 \pi \sigma
a^{2} {(1+{e^{2}\over{2}})}^{2} }} \right] }^{1/4}, \eqno(8) $$
where the quantity $a(1+e^{2}/2)$ is the time-averaged distance between
the planet and the star for a Keplerian orbit of semi-major axis
$a$ and eccentricity $e$, and $A$ is the Bond albedo for the planet.
We adopt $A=0.4$ when computing the values for $T_{\rm eq}$ listed
in Table 4.

The last eight rows of Table 4 show our predictions for the planetary
radii and circularization e-folding times. The radii are computed via
linear interpolation between the relevant models of Tables 1 and 2,
and are coded according to the presence or absence of a core (c/nc)
and the presence or absence of kinetic heating (k/nk). Circularization
times in each case are computed assuming a fiducial value
$Q_{\rm P} =10^{6}$, and $\sin i =1.0$. Note that $\tau_{\rm circ}$ is 
linearly dependent on $Q_{\rm P}$.

With these assumptions, we find that for HD 68988b and HD 217107b, the
observationally well-established eccentricities of $e=0.14$ for both
planets are consistent with computed circularization timescales that
comfortably exceed the estimated stellar ages. In particular, we note that
the non-zero observed eccentricity for HD 68988b is no longer in
disagreement with the circularization timescale. A disagreement
was noted previously by Vogt et al. (2002), who computed a $\tau_{\rm circ}=
1.8~{\rm Gyr}$ based on a radius estimate for HD 68988b of $R=1.4 \, R_{\rm JUP}$, 
motivated by the observed radius for HD 209458b.

We also note that $\dot E_d = 2 \times 10^{25}$ and $1 \times 10^{25} (10^6/Q_{\rm P}) 
(R_{\rm P}/R_{\rm JUP})^5\,$ erg s$^{-1}$ for HD 68988b and HD 217107b respectively.  
In contrast, the assumed kinetic energy deposition rate (1\% of the
irradiation flux) is $\dot E_k = 5 \times 10^{26} (R_{\rm P}/R_{\rm JUP})^2\,$erg 
s$^{-1}$ for both planets.  Even with a modest Q-value ($2.5 \times 
10^5$), $\dot E_d$ is an order of magnitude smaller than $\dot E_k$
so that tidal dissipation is unlikely to enlarge these planets by a
noticeable amount.

For HD 168746, however, the situation is more interesting. Because the
planet mass is likely to be small, $M~\sin i  = 0.23~M_{\rm JUP}$, there
is a large range in the predicted planetary radii depending on the presence
of a core and/or kinetic heating, and hence the estimates for $\tau_{\rm circ}$
vary widely, from $\tau_{\rm circ}=0.26 \, {\rm Gyr}$ for a core-free model
with kinetic heating, to $\tau_{\rm circ}=3.5 \,{\rm Gyr}$ for a model with a
core and no kinetic heating. Pepe et al. (2002) estimate that the star
HD 168746 is at least several Gyr old, and report an observed eccentricity of
$e=0.081\pm0.02$. Using the 154 RVs posted at CDS in conjunction
with the Pepe et al. (2002) paper, we confirm that a best-fit $\sqrt {\chi^{2}}=
1.51$ is obtained for $e=0.081$, but this value increases to only
$\sqrt {\chi^{2}}=1.54$ for $e=0$, indicating that a circular orbit
for HD 168746b is still tenable. Improvement of the orbital elements will
certainly result as further RV measurements are accumulated.

For this low-mass planet, $\dot E_d = 6 \times 10^{24} (10^6/Q_{\rm P})
(R_{\rm P}/R_{\rm JUP})^5\,$erg s$^{-1}$.  We computed two sets (with and 
without a core of 20 $M_{\oplus}$) of self-consistent
models in which uniform tidal dissipation (in accordance with an
eccentricity $e=0.081$) is applied to the planetary
interior, with $Q_{\rm P} = 10^6$, in addition
to the surface irradiation (Bodenheimer et al. 2001). 
Since the radius as a function of time is not known in advance,
we iterate the solution until $\dot E_d$ at the end of the calculation
matches the computed model luminosity.
For models with and without cores, $R_{\rm P} =1.01 \, R_{\rm JUP}$ and $1.33 \, R_{\rm JUP}$
respectively.  These tidal models fall between the dissipationless and
{\it ad hoc} kinetic energy dissipation models in the cases both with and
without cores.  Although the coreless tidal dissipation model gives the
same value for $R_{\rm P}$ as the core-structure model with kinetic energy
input, the observational determination of $e$ will remove the degeneracy.

Finally, we remark on the recently reported planet
OGLE-TR-56b,
$M_p = 0.9 \, M_{\rm JUP}$, $P=1.2 \, {\rm d}$,
$T_{\rm eq} = \, 1,900$ K (Konacki  et al. 2003).  In
comparison with the results in Table 1, the observationally measured
$R_{\rm P} = 1.3 \pm 0.15 \, R_{\rm JUP}$ is slightly larger than that determined for
the irradiated planet with or without a core.  But the results in
Table 2 indicate that kinetic heating at the rate of $1.7\%$ of the surface
irradiative flux induces the planet, with or without a core, to have
$R_{\rm P}$ larger than the observed value.  Thus, kinetic heating, if it
occurs below the surface of a short-period planet, is less efficient than
assumed by Guillot \& Showman (2002).  We also note that in order to
prevent tidal dissipation from expanding $R_{\rm P}$ beyond
its observed value,
the eccentricity of OGLE-TR-56b must be less than $1-2 \times 10^{-3}$,
depending on its $Q_{\rm P}$ value.  If the observed parameters are
confirmed, the eccentricity damping time scale of the planet $\tau_{\rm circ}
\simeq 1.8 (Q_{\rm P}/10^6)$ Myr would be much shorter than the main sequence life
span of the host star.

\section{Discussion}

Our evolutionary models for giant planets indicate that the planetary
radius is sensitively dependent on both the presence or absence of a core
and on the amount of energy that is deposited within the interior of the
planet. This dependence is strongest for hot low-mass (i.e. $M< 1~M_{\rm JUP}$)
planets, while being quite weak for planets with large mass
(i.e. $M \approx  3~M_{\rm JUP}$) regardless of the equilibrium  surface temperature.
The discovery of additional extrasolar planets which transit bright
parent stars will impart a great deal of information on the structure
and evolution of giant planets in general. The number of such planets,
however, will be quite limited. The parent star of HD 209458b has $V=7.65$,
yet photometric measurements with HST are photon-limited to a cadence
that resolves the critical ingress and egress periods into 80 second time
bins (Brown et al. 2001), and a large number of additional RV measurements
will be required in order to determine whether the eccentricity of HD 209458b
is being forced by a second companion. A similar follow-up effort will
be required for additional transiting planets as they are found.

The number of available transiting planets with $3~{\rm d}< P < 200~{\rm d}$
is severely limited by transit probabilities that are generally less than
10\%. Information on the interior properties of short-period planets
can nonetheless be obtained by examining which of the planetary orbits
have been tidally circularized. Circularization times scale as the fifth
power of the planetary radius. As a large aggregate of planets is built
up, increasingly stringent limits can be placed on the unknown tidal quality
factor $Q_{\rm {P}}$ and the planetary radii. 

Finally, we note that the discovery of intermediate-period transiting
planets is a challenging observational task. The most cost-effective way
to find these transiting planets is to harness a network of small independent
telescopes to obtain multiple (i.e. 3-5 separate observers) differential-
photometric time-series of known planet-bearing stars during the well-defined
time windows in which transits are expected to occur. We are currently
pursuing this strategy using a network of amateur observers and small-college
observatories.\footnote{http://www.transitsearch.org}

We thank Richard Freedman for providing us with Rosseland mean opacities
for our planetary models.
We thank Geoff Marcy for providing us with the latest Keck RV data
set for HD 209458 in advance of publication.
We thank Debra Fischer, Richard Freedman, Geoff
Marcy and Mark Marley for useful conversations.
This work was supported in part by a NASA
astrophysics theory program which supports a joint Center for Star
Formation Studies at NASA-Ames Research Center, UC Berkeley and UC
Santa Cruz, by the NASA Origins program through grant 
NCC2-5501, and by the NSF through grant AST-9987417.

\clearpage

\begin{figure}
\caption{
Distribution of 98 planetary periods, $M\sin i $'s, and parent star magnitudes
versus date of discovery announcement.
}
\end{figure}
\clearpage

\begin{figure}
\caption{
Power spectrum of residuals to the best 1-planet fit to HD 209458. 
}
\end{figure}
\clearpage

\begin{figure}
\caption{
Secular exchange of eccentricity in a 2-planet model of the
HD 209458b data set. Time is given in years. The solid curves
show the eccentricities of planet ``b'' and the hypothetical
planet ``c''. The dotted lines show the eccentricity
exchange which would occur in the absence of relativistic
precession.
}
\end{figure}
\clearpage

\newpage
\bigskip
\centerline{\bf \sc Table 1: Predicted Radii of Irradiated Giant Planets at T=4.5 Gyr}
\medskip

\begin{center}
\begin{tabular}{lllllllllll}
\hline
\hline
& $0.11~{\rm M_{J}}$ & & $0.23~{\rm M_{J}}$ & & $0.69~{\rm M_{J}}$ & & $1.0~{\rm M_{J}}$ & & $3.0~{\rm M_{J}}$ &  \\
\hline
$T_{eq}$&core&no core&core&no core&core&no core&core&no core&core&no core\\
\hline
2000&0.87&1.75&1.07&1.31&1.10&1.22&1.14&1.20&1.18&1.19\\
1500&0.74&1.20&0.95&1.14&1.03&1.13&1.08&1.13&1.12&1.13\\
1000&0.69&1.09&0.90&1.07&1.01&1.10&1.06&1.11&1.11&1.11\\
 500&0.66&1.01&0.88&1.03&1.00&1.09&1.05&1.10&1.10&1.11\\
 113&0.61&0.89&0.82&0.95&0.95&1.03&1.01&1.05&1.07&1.07\\
\hline
\hline
\end{tabular}
\end{center}

\bigskip
\centerline{\bf \sc Table 2: Predicted Radii of Irradiated Giant Planets at T=4.5 Gyr}
\centerline{\bf \sc (Models With Kinetic Heating)}
\medskip

\begin{center}
\begin{tabular}{lllllllllll}
\hline
\hline
& $0.11~{\rm M_{J}}$ & & $0.23~{\rm M_{J}}$ & & $0.69~{\rm M_{J}}$ & & $1.0~{\rm M_{J}}$ & & $3.0~{\rm M_{J}}$ &  \\
\hline
$T_{eq}$&core&no core&core&no core&core&no core&core&no core&core&no core\\
\hline
2000&$>$2.0&$>$2.0&$>$2.0&$>$2.0&1.74&1.81&1.47&1.61&1.40&1.35\\
1500&1.69&$>$2.0&1.61&1.80&1.35&1.51&1.30&1.38&1.25&1.26\\
1000&1.14&1.72  &1.34&1.53&1.16&1.27&1.13&1.18&1.13&1.14\\
 500&0.80&1.13  &1.02&1.14&1.02&1.11&1.06&1.11&1.10&1.11\\
 113&0.61&0.89  &0.82&0.95&0.95&1.03&1.01&1.05&1.07&1.07\\
\hline
\hline
\end{tabular}
\end{center}
\bigskip

\newpage

\centerline{\bf \sc Table 3: Trial 2-Planet Fits to HD 209458 data}
\medskip
\begin{center}
\begin{tabular}{lllllllll}
\hline
\hline
&Fit 1& &Fit 2 & &Fit 3& &Fit 4 &\\
Parameter & b & c & b & c & b & c & b & c\\
\hline
$P$ (d) & 3.5248(f) & 84.17 & 3.5248(f) & 84.37 & 3.5248(f) & 84.29 & 3.5248(f)& 84.714\\
$M$ (deg) & 221.85(f) & 91.59 & 221.85(f) & 101.22 & 221.85(f) & 104.96 & 241.57 & 154.48\\
$e$ & 0.025 & 0.00(f) & 0.022 & 0.20(f)&0.019&0.40(f)& 0.00037&0.697\\
$\varpi$ & 67.58 & 60.5(f)&67.75& 60.5(f)&67.74&60.5(f)& 47.6&31.2\\
$M_{\rm JUP}\sin(i)$ & 0.679 & 0.127(f)&0.679&0.127(f)&0.679&0.127(f)&0.64 &0.227\\
\hline
\hline
\end{tabular}
\end{center}
\newpage
{}
\bigskip
\centerline{\bf \sc Table 4: Planets with $6{\rm d}<P<8{\rm d}$}
\medskip
\begin{center}
\begin{tabular}{llll}
\hline
\hline
 & HD 68988b & HD 168746 & HD 217107\\
\hline
$P$ (d) & 6.276 & 6.403 & 7.125\\
$K$ (m/s) & 187 & 28 & 139.7\\
$M \sin(i) (M_{\rm JUP})$ & 1.80 & 0.23 & 1.29\\
$e$ & 0.14 & 0.081 & 0.14\\
$M_{\star}/M_{\odot}$ & 1.11 & 1.04 & 0.98\\
${T_{\rm eff}}_{\star}$ (K) & 5888 & 5754 & 5623\\
${T_{\rm eq,P}}$ (K) & 1004 & 973  & 919 \\
$R_{\star}/R_{\odot}$ & 1.17 & 1.07 & 1.12 \\
Stellar Age (Gyr) & 6 & $>2 \,$ Gyr & 5.6 \\
$R_{P}({\rm  c,nk})/R_{\rm JUP}$ & 1.08 & 0.90 & 1.07 \\
$R_{P}({\rm nc,nk})/R_{\rm JUP}$ & 1.11 & 1.07 & 1.11 \\
$R_{P}({\rm  c, k})/R_{\rm JUP}$ & 1.13 & 1.34 & 1.12 \\
$R_{P}({\rm nc, k})/R_{\rm JUP}$ & 1.17 & 1.53 & 1.16 \\
$\tau_{\rm circ}({\rm  c,nk})$ (Gyr) & 11.5 & 3.5  & 13.9 \\
$\tau_{\rm circ}({\rm nc,nk})$ (Gyr) & 10.1 & 1.5  & 11.6 \\
$\tau_{\rm circ}({\rm  c, k})$ (Gyr) &  9.2 & 0.51 & 10.8 \\
$\tau_{\rm circ}({\rm nc, k})$ (Gyr) &  7.9 & 0.26 &  9.1 \\
\hline
\hline
\end{tabular}
\end{center}

{}


\begin{thebibliography}{DUM}

\bibitem[]{}
Alexander, D. R., \& Ferguson, J. W. 1994, ApJ, 437, 879 

\bibitem[]{}
Allende Prieto, C., \& Lambert, D. L. 1999, A\&A, 352, 555

\bibitem[]{}
Barnes, J. W., \& O'Brien, D. P. 2002, ApJ, 575, 1087

\bibitem[]{}
Bodenheimer, P., Hubickyj, O., \& Lissauer, J. J. 2000, Icarus, 143, 2

\bibitem[]{}
Bodenheimer, P., Lin, D. N. C., \& Mardling, R. A. 2001, ApJ, 548, 466

\bibitem[]{}
Brown, T. M., Charbonneau, D., Gilliland, R. L., Noyes, R. W., \& Burrows, A.
2001, ApJ, 552, 699

\bibitem[]{}
Bundy, K. A., \& Marcy, G. W. 2000, PASP, 112, 1421

\bibitem[]{}
Burrows, A., Guillot, T., Hubbard, W. B., Marley, M. S., Saumon, D., Lunine, J. I.,
\& Sudarsky, D. 2000, ApJ, 534, L97

\bibitem[]{}
Burrows, A., Hubbard, W. B., Lunine, J. I., \& Liebert, J. 2001, Rev. Mod. Phys., 
73, 719 

\bibitem[]{}
Castellano, T., Jenkins, J., Trilling, D. E., Doyle, L., \& Koch, D.
2000,  ApJ, 532, L51

\bibitem[]{}
Chabrier, G., \& Baraffe, I. 2000, ARA\&A, 38, 337 

\bibitem[]{}
Charbonneau, D., Brown, T. M., Latham, D. W., \& Mayor, M. 2000, ApJ, 529, L45

\bibitem[]{}
Charbonneau, D., Brown, T. M., Noyes, R. W., \& Gilliland, R. L. 2002, ApJ,
568, 377

\bibitem[]{}
Cody, A. M., \& Sasselov, D. D. 2002, ApJ, 569, 451

\bibitem[]{}
Dobbs-Dixon, I., Lin, D. N. C., \& Mardling, R. A. 2003, ApJ, submitted

\bibitem[]{}
Evonuk, M., Lin, D. N. C., \& Mardling, R. A. 2003, ApJ, submitted

\bibitem[]{}
Fischer, D. A., Marcy, G. W., Butler, R. P., Vogt, S. S., \& Apps, K. 1999,
PASP, 111, 50

\bibitem[]{}
Fischer, D. A., Marcy, G. W., Butler, R. P., Vogt, S. S., Frink, S. F., \&
Apps, K. 2001,  ApJ, 551, 1107

\bibitem[]{}
Goldreich, P., \& Soter, S. 1966, Icarus, 5, 375

\bibitem[]{}
Gu, P. G, Lin, D. N. C., \& Bodenheimer, P. 2003, ApJ, in press

\bibitem[]{}
Guillot, T., Burrows, A., Hubbard, W. B., Lunine, J. I., \& Saumon, D. 1996,
ApJ, 459, L35

\bibitem[]{}
Guillot, T., \& Showman, A. P. 2002, A\&A, 385, 156

\bibitem[]{}
Henry, G. W., Marcy, G. W., Butler, R. P., \& Vogt, S. S. 2000, ApJ, 529, L41

\bibitem[]{}
Henyey, L. G., Forbes, J. E., \& Gould, N. L. 1964, ApJ, 139, 306

\bibitem[]{}
Hubbard, W. B., Burrows, A., \& Lunine, J. I. 2002, ARA\&A, 40, 103 

\bibitem[]{}
Konacki, M., Torres, G., Jha, S. \& Sasselov D. D. 2003 Nature,
421, 507

\bibitem[]{}
Laughlin, G., \& Chambers, J. E. 2001, ApJ, 551, L109  

\bibitem[]{}
Lin, D. N. C., Bodenheimer, P., \& Richardson, D. C. 1996, Nature, 380, 606



\bibitem[]{}
Mardling, R. A., \& Lin, D. N. C. 2002, ApJ, 573, 829

\bibitem[]{}
Marley, M. S. 1999. In Encyclopedia of the Solar System, ed.
P. R. Weissman, L. McFadden, \& T. V. Johnson (San Diego: Academic
Press), p. 339

\bibitem[]{}
Mayor, M., \& Queloz, D. 1995, 
Nature, 378, 355 

\bibitem[]{}
Mazeh, T., Naef, D., Torres, G., Latham, D. W., Mayor, M. M., Beuzit, J.-L.,
Brown, T. M., Buchhave, L., Burnet, M., Carney, B. W., Charbonneau, D.,
Drukier, G. A., Laird, J. B., Pepe, F., Perrier, C., Queloz, D., Santos, N. C.,
Sivan, J.-P., Udry, S., \& Zucker, S. 2000, ApJ, 532, L55

\bibitem[]{}
Miralda-Escud\'{e}, J. 2002, ApJ, 564, 1019

\bibitem[]{}
Murray, C. D.,  \& Dermott, S. F. 1999, Solar System Dynamics
(Cambridge: Cambridge University Press).

\bibitem[]{}
Pepe, F., Mayor, M., Galland, F., Naef, D., Queloz, D., Santos, N. C.,
Udry, S., \& Burnet, M. 2002, A\&A, 388, 632

\bibitem[]{}
Pollack, J. B., Hubickyj, O., Bodenheimer, P., Lissauer, J. J., Podolak, M.,
\& Greenzweig, Y. 1996, Icarus, 124, 62

\bibitem[]{}
Press, W. H., Teukolsky, S. A., Vetterling, W. T., \& Flannery, B. P. 1992,
 Numerical Recipes: The Art of Scientific 
Computing, 2nd Edition  (Cambridge: Cambridge Univ. Press)  

\bibitem[]{}
Queloz, D., Eggenberger, A., Mayor, M., Perrier, C., Beuzit, J. L.,
Naef, D., Sivan, J. P., \&  Udry, S. 2000, A\&A, 359, L13

\bibitem[]{}
Robichon, N., \& Arenou, F. 2000, A\&A, 355, 295

\bibitem[]{}
Saar, S. H., Butler, R. P., \& Marcy, G. W. 1998, ApJ, 498, L153

\bibitem[]{}
Saumon, D., Chabrier, G., \& van Horn, H. M. 1995, ApJS, 99, 713 

\bibitem[]{}
Scargle, J. D. 1982, ApJ, 263, 835

\bibitem[]{}
Terquem, C., Papaloizou, J. C. B., Nelson, R. P., \& Lin, D. N. C. 1998, 
ApJ, 502, 788

\bibitem[]{}
Tsuji, T. 2002, ApJ, 575, 264 

\bibitem[]{}
Vogt, S. S., Butler, R. P., Marcy, G. W., Fischer, D. A., Pourbaix, D.,
Apps, K., \& Laughlin, G. 2002, ApJ, 568, 352

\bibitem[]{}
Welsh, W. F., et al. 2003, preprint

\bibitem[]{}
Wuchterl, G., Guillot, T., \& Lissauer, J. J. 2000, In Protostars
and Planets IV, ed. V. Mannings, A. P. Boss, \& S. S. Russell
(Tucson: University of Arizona Press), p. 1081

\bibitem[]{}
Yoder, C. F., \& Peale, S. J. 1981, Icarus, 47, 1


\end{thebibliography}
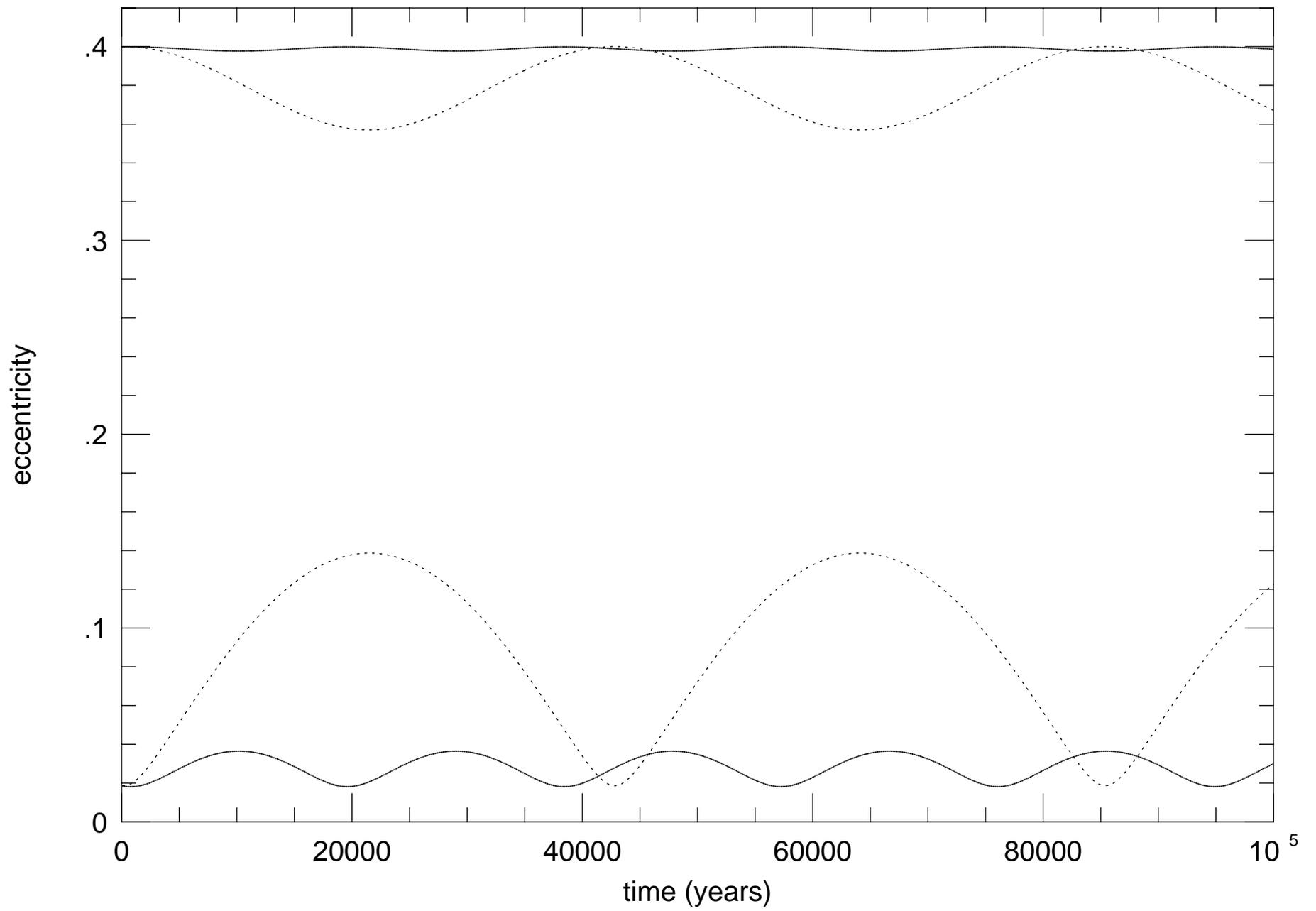
\end{document}